\documentclass[prl,amsmath,amssymb,twocolumn,superscriptaddress,showpacs]{revtex4-1}
\usepackage{bm}
\usepackage{amssymb}
\usepackage{colordvi}
\usepackage{graphicx}
\usepackage{color}
\usepackage{hyperref}
\newcommand{\up}{\uparrow}
\newcommand{\down}{\downarrow}

\newcommand{\be}{\begin{equation}}
\newcommand{\ee}{\end{equation}}
\newcommand{\bea}{\begin{eqnarray}}
\newcommand{\eea}{\end{eqnarray}}
\newcommand{\ep}{\epsilon}
\newcommand{\vep}{\varepsilon}

\newcommand{\ome}{\omega}
\newcommand{\Ome}{\Omega}

\def\nn{\nonumber}

\def\ket#1{\vert #1 \rangle}

\def\grad {\mbox{\boldmath$\nabla$\unboldmath}}

\begin{document}

\title{Possible Realization of Optical Quadratic and Dirac Points in Woodpile Photonic Crystals}

\author{Hai-Xiao Wang}
\affiliation{School of Physical Science and Technology, Guangxi Normal University, Guilin 541004, China}
\affiliation{School of Physical Science and Technology, \&
	Collaborative Innovation Center of Suzhou Nano Science and
	Technology, Soochow University, 1 Shizi Street, Suzhou 215006, China}
\affiliation{Department of Physics and Center for Theoretical Physics, National Taiwan University, Teipei 10617, Taiwan}
\author{Yige Chen}
\affiliation{Department of Physics, University of Toronto, Toronto, M5S 1A7, Canada}
\author{Guang-Yu Guo}\email{gyguo@phys.ntu.edu.tw}
\affiliation{Department of Physics and Center for Theoretical Physics, National Taiwan University, Teipei 10617, Taiwan}
\affiliation{Physics Division, National Center for Theoretical Sciences, Hsinchu 30013, Taiwan}
\author{Hae-Young Kee}\email{hykee@physics.utoronto.ca}
\affiliation{Department of Physics, University of Toronto, Toronto, M5S 1A7, Canada}
\affiliation{Canadian Institute for Advanced Research, Toronto, Ontario, M5G 1Z8, Canada}
\author{Jian-Hua Jiang}\email{jianhuajiang@suda.edu.cn}
\affiliation{School of Physical Science and Technology, \&
  Collaborative Innovation Center of Suzhou Nano Science and
  Technology, Soochow University, 1 Shizi Street, Suzhou 215006,
  China}

\begin{abstract}
The simulation of fermionic relativistic physics, e.g., Dirac and Weyl physics, has led to the discovery of many unprecedented phenomena in photonics, of which the optical-frequency realization is, however, still challenging. Here, surprisingly, we discover that the woodpile photonic crystals commonly used for optical frequency applications host exotic fermion-like relativistic degeneracies: a Dirac nodal line and a fourfold quadratic point, as protected by the nonsymmorphic crystalline symmetry. Deforming the woodpile photonic crystal leads to the emergence of type-II Dirac points from the fourfold quadratic point. Such type-II Dirac points can be detected by its anomalous refraction property which is manifested as a giant birefringence in a slab setup. Our findings provide a promising route towards 3D optical Dirac physics in all-dielectric photonic crystals. 
\end{abstract}

\maketitle

{\it Introduction.}--- 
Fermionic relativistic waves described by the Dirac and Weyl equations~\cite{volovik,wan,fang,liu,WPII} (and beyond~\cite{science}) have many nontrivial properties as discovered in condensed materials~\cite{mele}. Recently, photonic crystals (PCs) became a versatile platform to simulate such relativistic waves~\cite{ling1,ling-exp,ct-exp,szhang1,szhang2,ctwood,3ddp,xiao,sWP,atwater,type2,ExptypeIWeyl,Rechtsman1,SZhang,Rechtsman2,Rechtsman3}. Photonic relativistic waves have been harnessed for a number of fundamental phenomena and applications such as {\em Zitterbewegung}~\cite{ZB}, pseudodiffusive transport~\cite{psd}, zero-index metamaterials~\cite{zim}, synthetic magnetic fields for photons~\cite{sMag}, and anomalous refraction~\cite{type2}. In particular, fermion-like relativistic (i.e., even-fold band degeneracy) waves are more appealing than the boson-like (i.e., odd-fold band degeneracy) counterparts, since they are closely related to photonic topological insulators~\cite{haldane,wangzhen,CI1,hafezi,FTI,z2meta,ctti,shvets,huxiao,lumh,oe1,3dti,3dwti}. At optical frequencies, the magneto-optical and bianisotropic effects of natural materials, which are often used to create photonic topological insulators, become negligible. One then has to turn to all-dielectric PCs as low-dissipation optical crystalline materials. Recent reports on the experimental observations of photonic Weyl points at the infrared and even optical frequency regimes~\cite{ExptypeIWeyl,Rechtsman1,Rechtsman2,Rechtsman3} indicate that three-dimensional (3D) dielectric PCs still hold the promise toward Weyl physics. However, realizing 3D optical Dirac points (DPs) as a potential pathway towards optical 3D topological insulators is more challenging, since in PCs the spin degeneracy of photons is broken. Space symmetry must be leveraged to simulate both the fermion-like Kramers degeneracy and the parity inversion~\cite{type2}. With the limited types of available optical-frequency 3D PCs in the current technology, it is unknown which can lead to 3D photonic DPs.

In this Letter, we illustrate a practical route towards 3D optical DPs: using woodpile-like PCs---a prototype 3D optical-frequency PCs that have been fabricated with high quality~\cite{wood1,wood2,wood3,wood4,noda,book} [Fig.~1(a)]. Surprisingly, we find that the woodpile PCs host two types of exotic band degeneracies in the lowest photonic bands: a Dirac nodal line and a fourfold quadratic point (FQP), as protected by the nonsymmorphic crystalline symmetry. Starting from the motherboard of the woodpile PCs, type-II DPs can be created by deforming the unit-cell geometry. Interestingly, we find that the type-II DPs exhibit anomalous birefringence. Such birefringence is maximized when the incident light excite exactly the type-II DPs. These findings provide a promising path towards optical-frequency Dirac physics and the potential realization of 3D optical topological insulators in all-dielectric PCs that are compatible with optoelectronic integration and nano-photonic applications~\cite{noda,phc-valley}.

\begin{figure}[htb]
  \includegraphics[width=8.6cm]{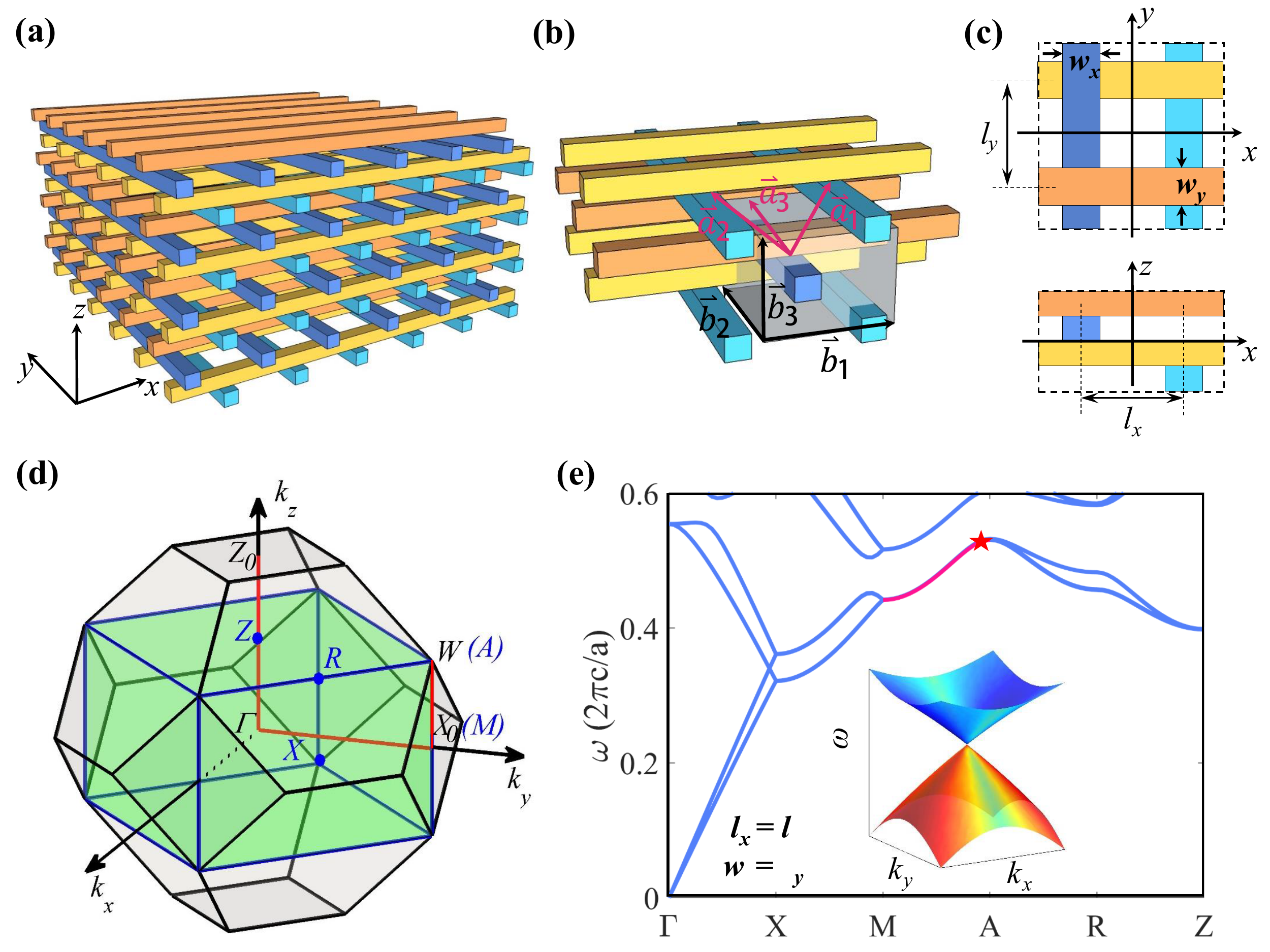}
  \caption{(Color online) (a) Optical-frequency woodpile PCs: layer-by-layer stacking of dielectric (colored) logs. (b) Lattice vectors of the undeformed woodpile PC, ${\vec a}_j$, are shown together with those of the deformed woodpile PC, ${\vec b}_j$ ($j=1,2,3$). (c) The top-view (upper) and the side-view (lower) of the unit-cell of the deformed (green) woodpile PC. (d) The relationship between the Brillouin zones of the undeformed (gray) and the deformed woodpile PCs. (e) The low-lying photonic bands of the undeformed woodpile PC with $l_x=l_y=b_x/2$, $w_x=w_y=0.4b_x$, $h=0.2b_x$, and $\vep=5.06$ (TiO$_2$). Inset: the Dirac dispersion at the $A$ point (labeled by the star). The $M$-$A$ line (red) is a Dirac nodal line on which each point has Dirac-like dispersions.}
\end{figure}

{\it Woodpile space symmetry.}--- Structures of the undeformed and deformed woodpile PCs are illustrated in Figs.~1(a)-(c) together with their unit-cells and lattice vectors. The two corresponding Brillouin zones are illustrated in Fig. 1(d). The deformed woodpile PC has a unit-cell twice as large as the unit-cell of the undeformed woodpile PC. These two PCs can be described in an unified fashion using the unit-cell of the deformed woodpile PC. In this study, we set the lattice constant as $|{\vec b}_1|=|{\vec b}_2|=b_x=0.8~\mu$m and $|{\vec b}_3|=b_z=0.8b_x$. The deformation of the woodpile PC can be feasibly realized by tuning the distances between the dielectric logs, $l_x$ and $l_y$, or by tuning the width of the logs, $w_x$ and $w_y$. The undeformed woodpile represents the limit with $w_x=w_y$ and $l_x=l_y=b_x/2$. We consider mostly the woodpile PCs made of TiO$_2$ which can be directly generalized to other dielectric materials such as silicon and GaAs.

The undeformed woodpile PC is a spiral stacking of the dielectric logs, with a four fold screw symmetry $S_{\frac{\pi}{2}}:=(x, y, z) \to (y, -x, z + \frac{b_z}{4})$. The space symmetry of the woodpile PCs is elaborated in details in the Supplemental Materials~\cite{SM}. We list here only the most relevant symmetries: $M_x := (x,y,z) \to (-x-\frac{b_x}{2}, y, z)$ and $M_y := (x,y,z) \to (x, -y-\frac{b_x}{2}, z)$, the glide symmetry $G_z := (x, y, z) \to (x + \frac{b_x}{2}, y, -z-\frac{b_z}{4})$, the fourfold screw symmetries $S_{\frac{\pi}{2}}$, and the point group symmetry $S_4 := (x,y,z) \to (y, - x, -z - \frac{b_z}{2})$. The deformed woodpile PCs may break most of the above symmetries, leaving some of the mirror or glide symmetries unchanged. Note that in this Letter, we use the capital letters for the symmetry operators while the small letters for their eigenvalues.

{\it Symmetry-enriched degeneracy.}---
Quite different from the simulation of Weyl points, the realization of the synthetic Kramers degeneracy ${\cal T}_p^2=-1$ for photons is crucial for the simulation of the Dirac and quadratic points as well as the Dirac nodal lines. 
Here, the synthetic Kramers degeneracy is realized via the nonsymmorphic crystalline symmetries. For
instance, the double degeneracy on the $k_z=\frac{\pi}{b_z}$ plane [the $A$-$R$-$Z$ line in Fig.~1(b)] can be understood by constructing an anti-unitary operator $\Theta_{\pi}\equiv S_{\frac{\pi}{2}}^2*{\cal T}$ which is invariant at the $k_z=\frac{\pi}{b_z}$ plane and yields $\left(\Theta_{\pi}\right)^2\Psi_{n,{\vec k}}=e^{ik_zb_z}\Psi_{n,{\vec k}}$ for {\em all} photonic Bloch states $\Psi_{n,{\vec k}}\equiv ({\vec E}_{n,{\vec k}}, {\vec H}_{n,{\vec k}})^T$ (including both the electric field ${\vec E}$ and the magnetic field ${\vec H}$ for the $n$-th band with a wavevector ${\vec k}$). Hence, for the $k_z=\pi/b_z$ plane
\be
\left.\Theta_\pi^2\right|_{k_z=\frac{\pi}{b_z}} = -1 
\ee
leads to the synthetic Kramers degeneracy for {\em all} photonic bands on the $k_z=\frac{\pi}{b_z}$
plane. Similarly, the double degeneracies on the $k_x=\frac{\pi}{b_x}$ [the $X$-$M$-$A$-$R$ line in Fig.~1(b)] and $k_y=\frac{\pi}{b_x}$ planes are induced by the other screw symmetries of the woodpile PCs (see
Supplemental Materials~\cite{SM} for details).

{\it Photonic Dirac nodal line.}--- The $M$-$A$ line is a nodal line composed of an ``infinite'' number of
two-dimensional (2D) DPs [see Fig.~1(e) and Fig.~2]. The two fundamental elements in the Dirac
equation, the spin and orbital degrees-of-freedom, are associated with the four degenerate states on the
line. The field profiles of these four modes [Fig.~2(a)] indicate that
they are the electric and magnetic dipole modes which can be denoted
by the mirror quantum number as $\ket{m_x,m_y}$ ($m_x,m_y=\pm 1$). The
fourfold degeneracy is dictated by $\Theta_z\equiv G_z*{\cal T}$
and $S_{\pi}\equiv S_{\frac{\pi}{2}}^2$ which are invariant operators on the $M$-$A$ line, as manifested by the following symmetric transformations (see Supplemental Materials for more details~\cite{SM}), 
\begin{subequations}
\begin{align}
& \Theta_z\ket{m_x,m_y} = \ket{-m_x,m_y}, \\
& S_{\pi}\ket{m_x,m_y} = \ket{-m_x,-m_y}, \\ 
& \Theta_zS_{\pi}\ket{m_x,m_y} = \ket{m_x,-m_y} .
\end{align}
\end{subequations}
Here, the ``orbital'' degree-of-freedom are associated with the {\em parity}, ${\cal P}=m_xm_y$. For example, the two
electric dipole modes $\ket{m_x=-1,m_y=1}$ and $\ket{m_x=1,m_y=-1}$ constitute the odd-parity
``antiparticle'' sector of the Dirac equation, whereas the magnetic dipole
modes $\ket{1,1}$ and $\ket{-1,-1}$ comprise the even-parity ``particle'' sector. The ``spin'' states for the particle sectors (`$p$') and antiparticle (`$a$') are constructed, respectively, as
\begin{align}
& \ket{p, \up} =
\frac{1}{\sqrt{2}}(\ket{1,1} + i \ket{-1,-1}) \equiv \frac{\ket{x^2-y^2}+i\ket{2xy}}{\sqrt{2}}, \nn \\
& \ket{p, \down} = \frac{1}{\sqrt{2}} (\ket{1,1} - i \ket{-1,-1}) \equiv \frac{\ket{x^2-y^2}-i\ket{2xy}}{\sqrt{2}}, \nn\\
& \ket{a, \up} =
\frac{1}{\sqrt{2}}(\ket{-1,1} + i \ket{1,-1}) \equiv \frac{\ket{x}+i\ket{y}}{\sqrt{2}},\nn \\
& \ket{a, \down} = \frac{1}{\sqrt{2}} (\ket{-1,1} - i \ket{1,-1}) \equiv \frac{\ket{x}-i\ket{y}}{\sqrt{2}}. 
\label{spin}
\end{align} 
In the above equations, we have denoted the
mirror eigenstates as $\ket{-1,1}\equiv \ket{x}$, $\ket{1,-1}\equiv
\ket{y}$,  $\ket{1,1}\equiv \ket{x^2-y^2}$ and $\ket{-1,-1}\equiv \ket{2xy}$ to elucidate the spatial symmetry of the eigenstates. We find that the two spin states carry finite and
opposite {\em total angular momenta} (including both spin and
orbital angular momentum) of photon (see Supplemental Materials~\cite{SM}), which is a natural generalization of the concept of emulating fermion-like spin
with photonic spin~\cite{z2meta} or orbital angular momentum~\cite{huxiao,oe1} in previous studies.

\begin{figure}[htb]
  \includegraphics[width=8.6cm]{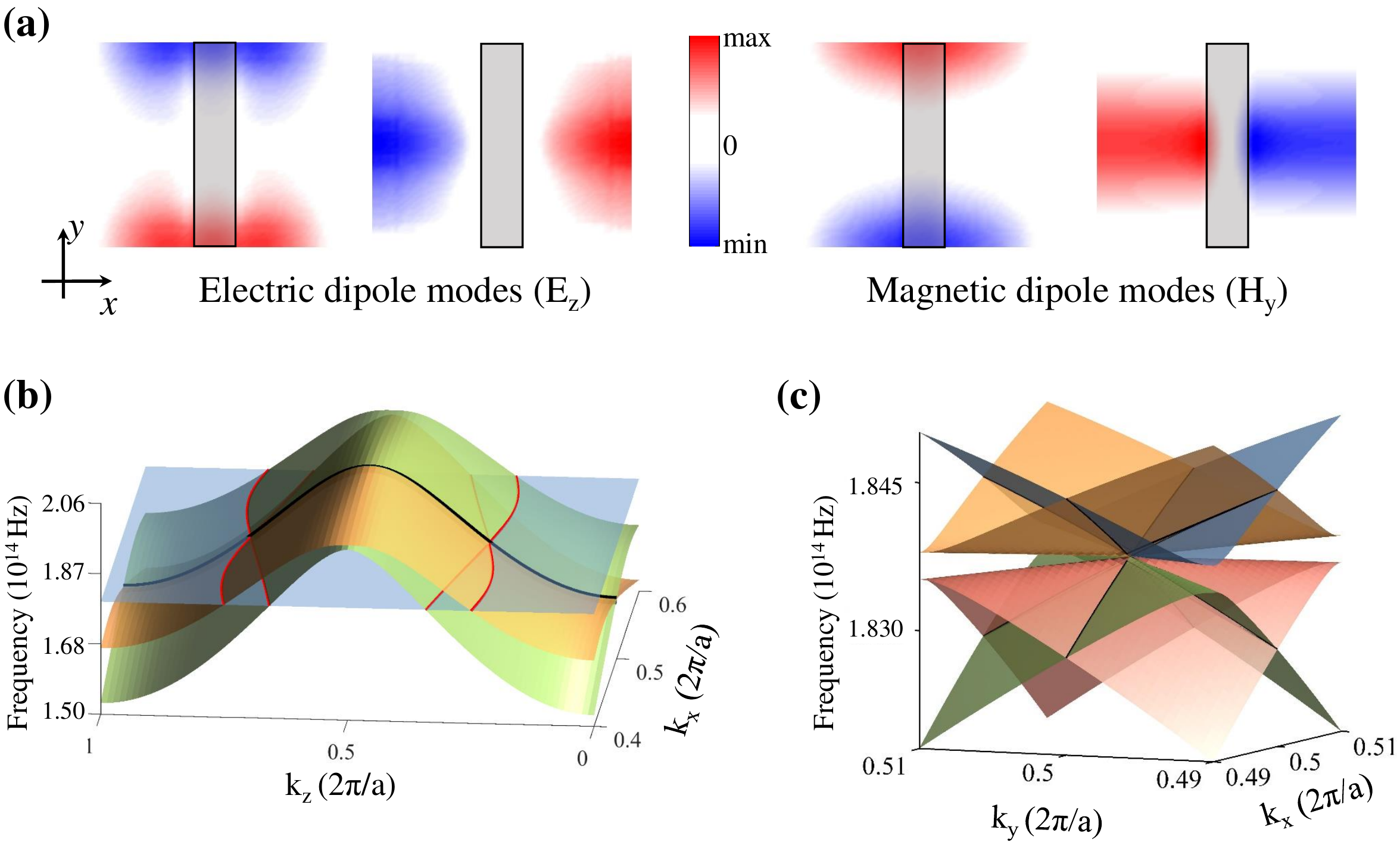}
  \caption{(Color online) (a) Field profiles of the four degenerate
    modes on the $M$-$A$ line. (b) and (c): Dispersion of the Dirac
    nodal line in the (b) $k_x$-$k_z$ and (c) $k_x$-$k_y$
    planes. In (b) an isofrequency plane (the blue-gray sheet) is
    plotted in order to show the isofrequency contours (the red
    curves). The Dirac nodal line is labeled by the black curve. Parameters are the same as in Fig.~1.}
\end{figure}

Photonic simulation of a fermionic Hamiltonian $\hat{{\cal H}}_F$ is the following mapping from the photonic Hamiltonian
$\hat{{\cal H}}_{EM}$ to the fermionic one (resembling the mapping
between the Dirac equation and the Klein-Gordon equation~\cite{Dirac}),
\be
\hat{{\cal H}}_{EM} := (\hat{{\cal H}}_F)^2 . \label{def}
\ee
From the Maxwell equation, $c^2\grad\times
\frac{1}{\vep}\grad\times{\vec H}=\ome^2{\vec H}$ ($c$ is the speed of light in vacuum, $\vep$ the relative
permittivity, and ${\vec H}$ is the magnetic field), the photonic
Hamiltonian can be defined as the Hermitian operator $\hat{{\cal
    H}}_{EM}:=c^2\grad\times \frac{1}{\vep}\grad\times$~\cite{book}. 

In the basis of the four states, $\ket{\rho,\up}$,
$\ket{\rho,\down}$ ($\rho=a,p$), the above mapping (together with the ${\vec k}\cdot{\vec P}$ theory) yields the following fermion-like Hamiltonian, 
\begin{align}
& \hat{{\cal H}}_F^{DL} = \ome_0 + v 
\left( \begin{array}{ccccc}
    0 & \hat{{\cal A}} \\
    \hat{{\cal A}}^\dagger &  0 \\
  \end{array}\right)  , \label{FDL}
\end{align}
Here $\hat{{\cal A}} \equiv \gamma_x q_x \sigma_x + \gamma_y q_y
\sigma_y$, and $q_j=k_j-\pi/b_x$ ($j=x,y$). The coefficients $\ome_0$, $v$,
$\gamma_x$ and $\gamma_y$ are $k_z$-dependent, making the Dirac nodal line. At the $M$ and $A$ points,
the $S_4$ symmetry imposes additional constraints, $\gamma_x=\gamma_y$,
leading to an isotropic 2D DP as shown in Fig.~1(e). Away from these
two points, the $S_4$ symmetry is ineffective, yielding unconventional 2D DPs as shown in
Fig.~2(c).

Since the particle and antiparticle sectors correspond to the magnetic and electric dipole modes, respectively, the
magneto-electric coupling in the Maxwell equations naturally guarantee
the linear in ${\vec q}$ ``spin-orbit couplings''. The coupling coefficients can be written as (Here, $j=x,y$, and UC stands for the unit cell)
\begin{align}
v\tilde{\gamma}_j & = \frac{c }{\sqrt{N_EN_H}}\int_{UC} d{\vec r}~ ({\vec E}_{\up, p}^\ast\times
  {\vec H}_{\down, a} + {\vec E}_{\down, a} \times {\vec H}_{\up,
    p}^\ast) \cdot {\vec n}_j, \nn 
\end{align}
where $\tilde{\gamma}_x=\gamma_x$ and
$\tilde{\gamma}_y=-i\gamma_y$; $N_E\equiv \int_{UC}d{\vec r}
\vep({\vec r})|{\vec E}|^2$,$N_H\equiv \int_{UC}d{\vec r} |{\vec
  H}|^2$; ${\vec n}_{j}$ is the unit vector along the $j=x,y$
direction; the integration is within a unit-cell. Interestingly, the
above expression is similar to the Poynting vector between the particle and antiparticle sectors. 
A general form of the Dirac velocity tensor is presented
in Supplemental Materials~\cite{SM} where the ``selection rules''
due to the mirror and glide symmetries are discussed.

\begin{figure}[htb]
  \includegraphics[width=8.6cm]{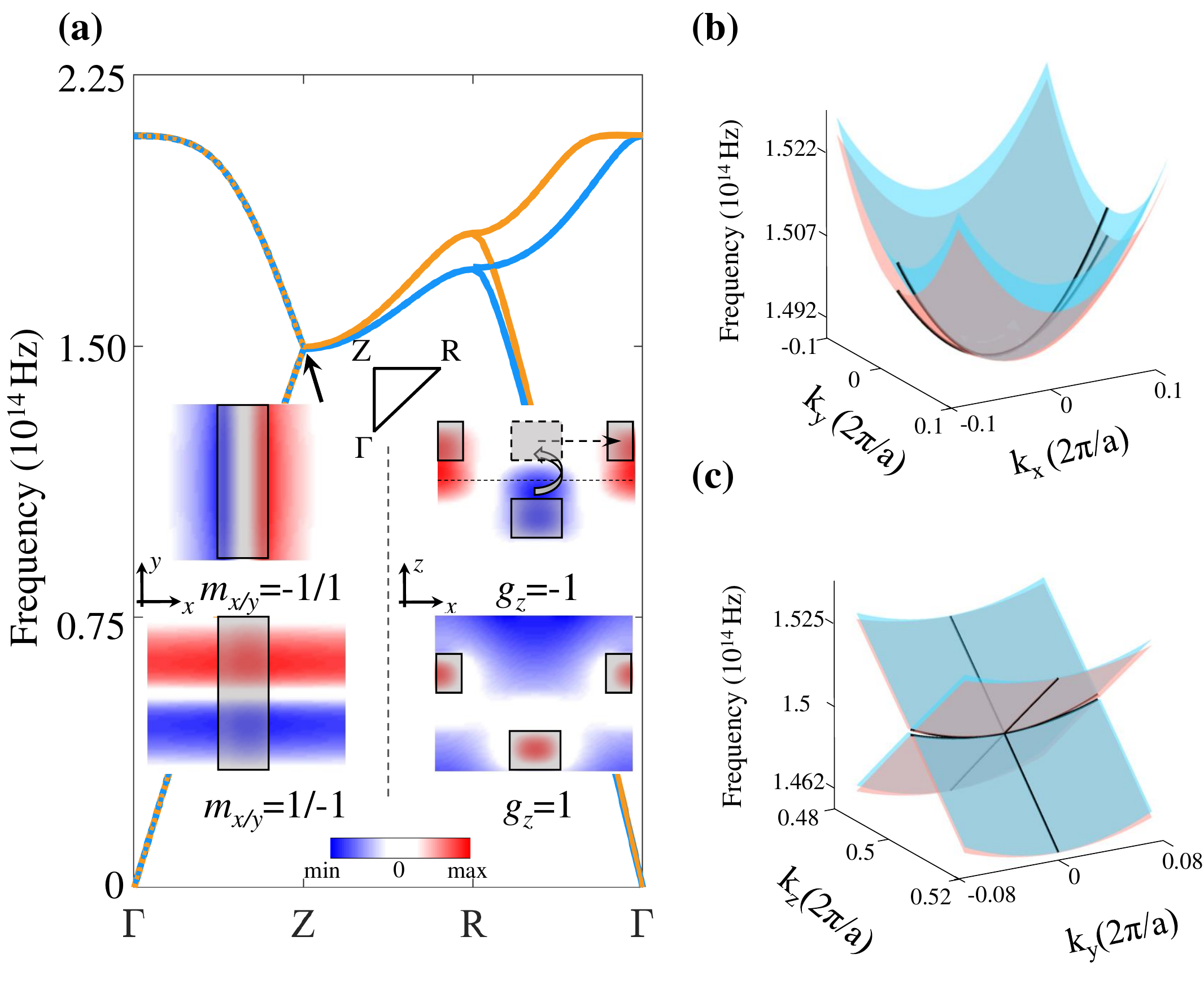}
  \caption{(Color online) Quadratic degeneracy points: (a) Photonic bands in the $k_y=0$ plane for
    the same parameters as in Fig.~1. The orange (blue) band has
    $m_y=+1$ (-1). The $Z$ point (indicated by the arrow) is a
    FQP with four eigenstates of different mirror
    ($m_{x/y}$) and glide ($g_z$) symmetries (illustrated in the
    inset). (b) and (c): Dispersion of the FQP in the (b) $k_x$-$k_y$ and (c) $k_y$-$k_z$
    planes.} 
\end{figure}

{\it Photonic fourfold quadratic point.}---
The $Z$ point is a photonic FQP which is induced by the fourfold screw symmetry $S_{\frac{\pi}{2}}$ as
\be
 (\Theta_{\frac{\pi}{2}})^4 =\left. e^{ik_zb_z}\right|_{k_z=\frac{\pi}{b_z}} = - 1 ,
\ee
where $\Theta_{\frac{\pi}{2}}\equiv S_{\frac{\pi}{2}}*{\cal T}$ transforms
$(k_x,k_y,k_z)$ to $(-k_y, k_x, -k_z)$ and is an invariant
operator at the $Z$ point. The above indicates that the quadruplet
consist of $\ket{\Psi}$, $\Theta_{\frac{\pi}{2}}\ket{\Psi}$,
$(\Theta_{\frac{\pi}{2}})^2\ket{\Psi}$, and
$(\Theta_{\frac{\pi}{2}})^3\ket{\Psi}$. If $\ket{\Psi}$ is labeled as
$\ket{m_x,m_y,g_z}$ ($g_z=\pm 1$ is the eigenvalue of $G_z$), then
\begin{align}
& \Theta_{\frac{\pi}{2}}\ket{\Psi}=\ket{m_y,m_x,g_z}, 
\ ~ (\Theta_{\frac{\pi}{2}})^2\ket{\Psi}=\ket{m_x,m_y,-g_z}, \nn\\
& (\Theta_{\frac{\pi}{2}})^3\ket{\Psi}=\ket{m_y,m_x,-g_z} .  \label{mglide}
\end{align}
The field patterns of the eigenstates are shown in Fig.~3(a), indicating $m_x=-m_y$. The fourfold degeneracy at the $Z$ point is protected by the screw symmetry $S_{\frac{\pi}{2}}$ and hence
independent of the specific parameters of the woodpile PC (see Supplemental
Materials~\cite{SM}).

The photonic system simulates the following fermion-like Hamiltonian of a 3D FQP, 
\begin{align}
& \hat{{\cal H}}_F^{Z} = \ome_Z + v_z q_z\hat{\tau}_y \nn\\
&\quad \quad \quad + f_0 [ (q_x^2-q_y^2) \hat{\sigma}_z + 2
f_1 q_x q_y \hat{\sigma}_x  + f_2 q_\parallel^2 ] , \label{Z}
\end{align}
where $\ome_Z$ is the frequency at the $Z$ point, $v_z$ is the group
velocity along the $z$ direction, and $q_z=k_z-\frac{\pi}{b_z}$. $\tau_z=\pm 1$
labels the $g_z=\pm 1$ states, while $\sigma_z=\pm 1$ labels the
$m_y=\pm 1$ states. The coefficients $f_i$ ($i=0,1,2$) depend on the
geometry and materials of the PC (see Supplemental
Materials~\cite{SM} for details).

{\it Photonic Dirac points.}--- There are two means to split the FQP to yield a pair of DPs: either tuning $w_y/w_x$ or $l_y/l_x$ away from unity. For these tuning, $S_{\frac{\pi}{2}}$ is broken while $(\Theta_{\frac{\pi}{2}})^2=S_{\pi}$ is preserved. From Eq.~(\ref{mglide}), the degeneracy between states of different $m_{x/y}$ is then lifted. This can be described by a constant perturbation $\Delta_zf_0 \sigma_z$ that splits the FQP to a pair of DPs emerging at two wavevectors ${\vec K}_{\pm} = (0, K_y^\pm, \frac{\pi}{b_z})$ with $K_y^\pm = \pm \sqrt{|\Delta_z|}$. The DPs are described by the following Hamiltonian,
\be
\hat{{\cal H}}_F^{DP\pm} = \ome_D \pm v_0 \delta k_y + v_z
\delta k_z\hat{\tau}_y \pm v_x \delta k_x \hat{\sigma}_x \pm v_y \delta k_y \sigma_z , 
\ee
where ${\vec \delta k}={\vec k}-{\vec K}_{\pm}$. Here, the coefficients are given by $\ome_D=\ome_Z+f_2f_0|\Delta_z|$, $v_0=2f_2f_0
K_y^+$, $v_x=2f_1f_0 K_y^+$, and $v_y=-2f_0K_y^+$.
With the parameters adopted, we find that $|v_0|>|v_y|$, thus the Dirac
cones are of the type-II nature [see Fig.~4(a)] (see Supplemental Materials~\cite{SM}).

\begin{figure}[htb]
  \includegraphics[width=8.6cm]{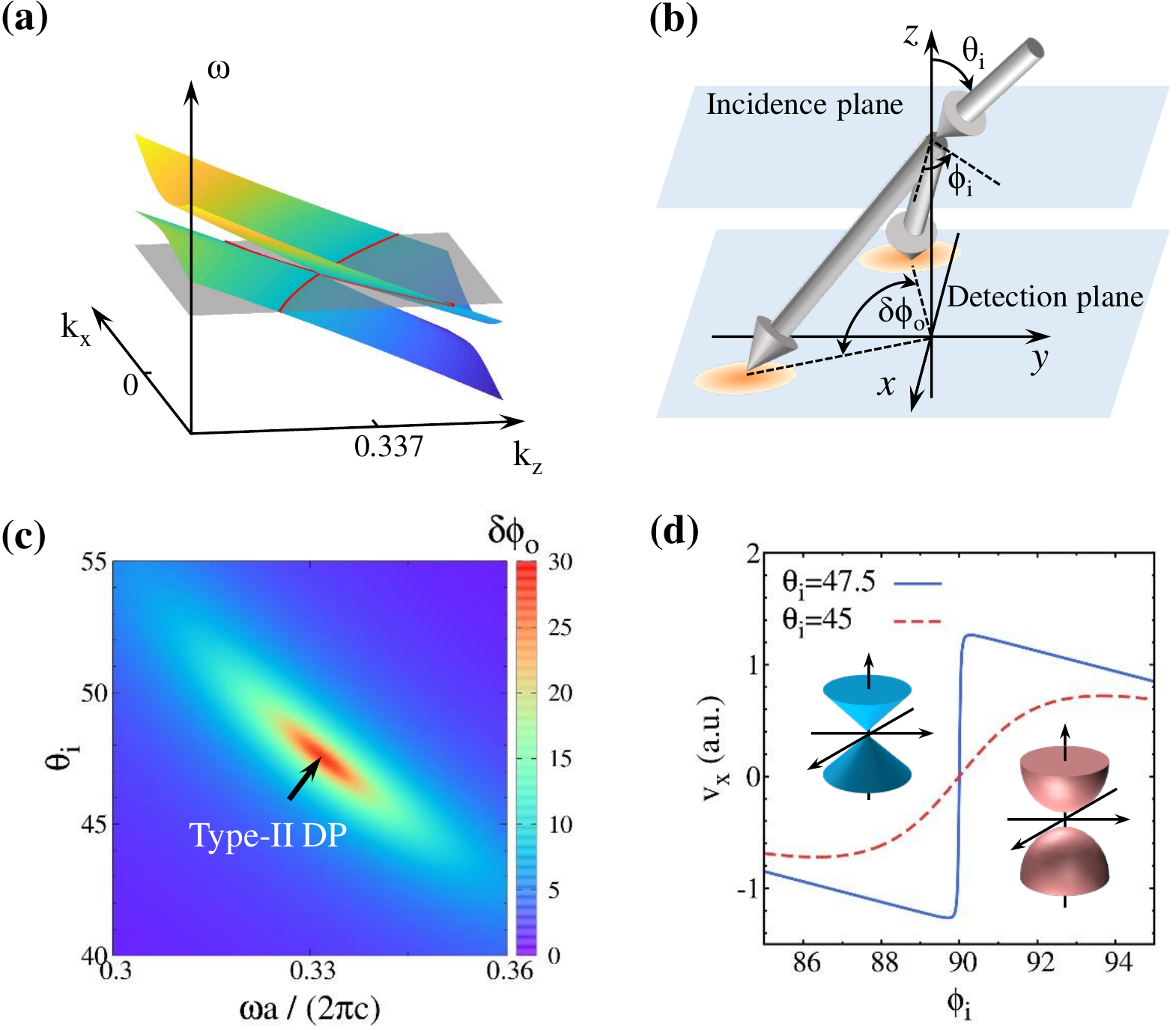}
  \caption{(Color online) (a) Photonic dispersion near a type-II DP
    for $l_x=l_y=0.5$, $w_x=0.2$, $w_y=0.3$, $h=0.25$, and
    $\vep=11$. (b) Illustration of the refraction angle
    measurement. $\theta_i$ and $\phi_i$ are angles of incidence,
    while $\delta\phi_o$ is the angular difference between the two
    refraction beams. (c) $\delta\phi_o$ vs. $\theta_i$ and $\ome$ at
    $\phi_i=89.5^\circ$ as a signature of the type-II DP. (d) The
    group velocity along $x$ direction $v_x$ for the beams correspond
    to the lower branch as a function of $\phi_i$ for $\theta_i=47.5$
    (aligned with the DP) and $\theta_i=45$ (misaligned with  the DP).}
\end{figure}

{\it Anomalous refraction.}---
The band degeneracies have been detected via transmission measurements~\cite{ling-exp,ExptypeIWeyl,Rechtsman1,SZhang,Rechtsman2,Rechtsman3}. Here, we show that they can also be detected via unconventional refraction, specifically, the birefringence. Birefringence emerges here because the type-II DPs support two branches of propagating refraction beams with different group velocities. The setup for measuring the birefringence is illustrated in
Fig.~4b. A Gaussian beam is shed on the PC slab and is detected at the
bottom of the slab. The drifted beam centers at the detection plane,
${\vec r}_o=r_{o}(\cos\phi_o,\sin\phi_o)$, gives the azimuth
refraction angle $\phi_o$. In the theory of refraction, the wavevector in
the PC is determined by frequency and parallel wavevector
matching. We find that the refraction angle is determined by the group
velocities as $\phi_o=\arctan(v_{g,y}/v_{g,x})$ and
$\theta_o=\arctan(\sqrt{v_{g,x}^2+v_{g,y}^2}/v_{g,z})$.

Considering refraction on the (001) surface, type-II dispersion yields
birefringence with two refraction beams~\cite{type2}, characterized by
two displacement vectors ${\vec r}_{o,\pm}$. We show that the DPs can
be detected by studying the difference in the azimuth
refraction angle $\delta\phi_o\equiv
|\phi_{o,+}-\phi_{o,-}|$. Fig.~4(c) shows that the DP at ${\vec 
  K}_+$ can be identified as the point with maximum $\delta\phi_o$
when frequency $\ome$ and angle of incidence $\theta_i$ are swept in large ranges at
a fixed azimuth incidence angle $\phi_i$. This phenomenon is a signature of the
type-II DPs: as the excited wavevector approaches the DP 
along the $k_x$ direction, the difference in $v_{g,x}$ for the upper
and lower branches changes abruptly. In contrast, away from the DP, such change is
gradual. Similar scenario happens when $\phi_i$ is swept at fixed
$\theta_i$ and $\ome$, as shown in Fig.~4(d). Therefore, the
difference in the azimuth refraction angle $\delta\phi_o$ is a signature of the DPs.

{\it Conclusion and outlook}.--- We have shown that the crystalline symmetry of woodpile PCs can enable the emergence of the 3D optical quadratic point and DPs. The DPs exhibit anomalous refraction and birefringence that can enable experimental detection of them. This study offers a guide for the realization of optical Dirac nodal line, FQP and DPs in 3D all-dielectric PCs. It also provides the inspiration and stimulation towards future realization of optical 3D topological insulators in all-dielectric PCs.

{\it Acknowledgments.}--- 
HXW and JHJ acknowledge supports from the Jiangsu provincial distinguished professor funding and the National Natural Science Foundation of China (Grant no. 11675116, 11904060, 12074281). JHJ also thanks Zhi Hong Hang, Jie Luo, Zhengyou Liu, and Huanyang Chen for many insightful discussions, as well as the University of Toronto and the Weizmann
Institute of Science for hospitality. YC and HYK are supported by NSERC of Canada Grant No. 06089-2016. HYK acknowledges support from the Canada Research Chairs program.

{}

\appendix

\begin{widetext}
\begin{center}
{\bf Supplemental Material for Optical Quadratic and Dirac Points in Woodpile Photonic Crystals}
\end{center}
\end{widetext}

\section{Sec.~A Connection between photonic bands in the tetragonal and face-centered cubic Brillouin zones}
Here, we illustrate the equivalence of photonic band in the scheme of tetragonal (TET) and face-centered-cubic (FCC) unit-cells under the case with $l_x=l_y=0.5$. For simplicity, we assume that the all lattice constants along the three directions are 1. As mentioned in the main text, the FCC unit-cell is the primitive unit-cell with only two logs, while the TET unit-cell has four logs. For the TET scheme, the lattice vectors are defined as follows:
\be
{\vec b}_1=(1,\ 0,\ 0), {\vec b}_2=(0,\ 1,\ 0), {\vec b}_3=(0,\ 0,\ 1).
\ee
While for the FCC scheme, we choose the lattice vexctors as follows:
\be
{\vec a}_1=\frac{1}{2}(-{\vec b}_1+{\vec b}_2+{\vec b}_3),
{\vec a}_2=\frac{1}{2}({\vec b}_1+{\vec b}_2+{\vec b}_3),
{\vec a}_3={\vec b}_2.
\ee
Thus $|{\vec a}_1|=|{\vec a}_2|=\frac{\sqrt{3}}{2}$,$|{\vec{a}_3}|=1$, and the 
volume of the FCC unit-cell is $\Ome=|\vec{a}_3 \cdot (\vec{a}_1 \times \vec{a}_2)|=\frac{1}{2}$.  It is known to all that the first Brillouin zone reduced by half when the unit-cell is doubled. In order to match the high symmetry points between the TET and FCC Brillouin zone, we need to rotate the unit-cell with 45 degree around $z-axis$, i.e., multiple the lattice vectors with rotation matrix $R$,
\be
\left(
\begin{matrix}
  \frac{1}{\sqrt{2}} & \frac{1}{\sqrt{2}} & 0 \\
  \frac{1}{\sqrt{2}} & -\frac{1}{\sqrt{2}} & 0 \\
  0 & 0 & 1 \\
\end{matrix}
\right)
\ee
Therefore, the new lattice vectors are:
\be
{\vec{a}_1}^{'}=(0,\ -\frac{\sqrt{2}}{2}, \frac{1}{2}),
{\vec{a}_2}^{'}=(\frac{\sqrt{2}}{2},\ 0,\ \frac{1}{2}),
{\vec{a}_3}^{'}=(\frac{\sqrt{2}}{2},\ -\frac{\sqrt{2}}{2},\ 0).
\ee
Finally, the new reciprocal lattice basis of FCC unit-cell can be derived as: 
\begin{align}
   &\vec{r}_{b1} =\frac{2\pi}{\Ome}({\vec{a}_2}^{'} \times {\vec{a}_3}^{'})
=2\pi (\frac{\sqrt{2}}{2}\ \frac{\sqrt{2}}{2}\ -1), \nn \\
   &\vec{r}_{b2} =\frac{2\pi}{\Ome}({\vec{a}_3}^{'} \times {\vec{a}_1}^{'})
=2\pi (-\frac{\sqrt{2}}{2}\ -\frac{\sqrt{2}}{2}\ -1),\nn \\
   &\vec{r}_{b3} =\frac{2\pi}{\Ome}({\vec{a}_1}^{'} \times {\vec{a}_2}^{'})
=2\pi (-\frac{\sqrt{2}}{2}\ \frac{\sqrt{2}}{2}\ 1)%
\end{align}
and the correspondence of the high symmetry points between FCC and TET Brillouin zone are list as follows: point $Z_0$, $X_0$, $W_0$ in the Brillouin zone under the FCC scheme refer to the point $\Gamma$, $M$, $A$ in the Brillouin zone under the TET scheme, respectively.
 
Fig. S1 shows the photonic bands below the fundamental gaps of woodpile in scheme of FCC and TET unit-cells, respectively. It should be noted there are four bands below the fundamental gap in the TET scheme, while there are only two bands in FCC scheme. In spite of the difference of the amount of the bands in two schemes, we remarked that the equivalence of these two schemes can be verified by band folding analysis. The fourfold degeneracy at the $Z$ point in the scheme of TET unit-cell originates from the band folding.

\begin{figure}[htb]
  \includegraphics[width=8.6cm]{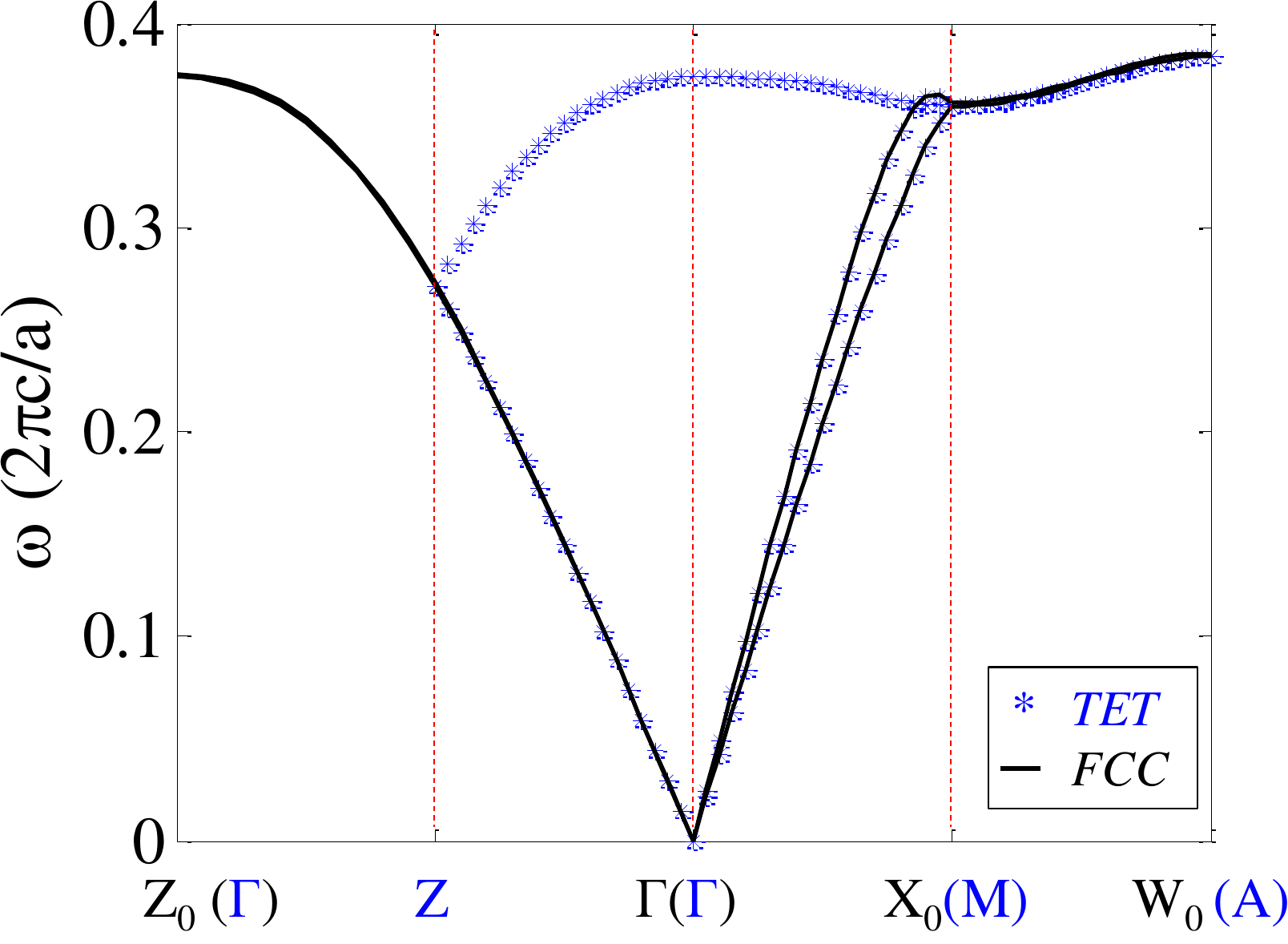}
  \caption{(Color online)  Correspondence of photonic bands of woodpile between two schemes: TET (blue-star) and FCC (black line) unit-cell, respectively. The parameter setting are list as follows: $l_x=l_y=0.5,w_x=w_y=0.25,\epsilon=13$. } 
\end{figure}

\section{Sec.~B Symmetries of Woodpile PCs}
The other important nonsymmorphic symmetries are the two-fold screw
symmetries: $S_{x} :
(x,y,z)\rightarrow\left(x+\frac{1}{2},-y+\frac{1}{2},-z-\frac{1}{4}\right)$ 
and $S_y : (x,y,z) \rightarrow
\left(-x+\frac{1}{2},y+\frac{1}{2},-z+\frac{1}{4}\right)$.
There are other two glide symmetries: $G_x :
(x, y, z) \to (-x, y + \frac{1}{2}, z + \frac{1}{2})$ and $G_y :
(x, y, z) \to (x + \frac{1}{2}, - y , z + \frac{1}{2})$.

\section{Sec.~C Other Kramers degeneracies}
The Kramers degeneracy on the $k_i=\pi (i=x,y)$ plane can be understood via the construction of the anti-unitary operators $\Theta_x\equiv S_x*{\cal T}: (x,y,z,
t)\rightarrow\left(x+\frac{1}{2},-y+\frac{1}{2},-z-\frac{1}{4},
  -t\right)$ and  $\Theta_y\equiv S_y*{\cal T}: (x,y,z,t) \rightarrow
  \left(-x+\frac{1}{2},y+\frac{1}{2},-z+\frac{1}{4}, -t\right)$, respectively. Let us consider the anti-unitary operator $\Theta_x$ as an example. $\Theta_x$ transforms ${\vec k}$ to $(-k_x, k_y, k_z)$ is a symmetry operator on the $k_x=\pi$ plane. The square of this operator is $\left(\Theta_{x}\right)^2: (x,y,z,t) \rightarrow (x+1,y,z,t)$, which yields $\left(\Theta_{x}\right)^2 \Psi_{n,{\vec k}} =e^{ik_x}\Psi_{n,{\vec k}}$ for {\em all} photonic Bloch states $\Psi_{n,{\vec k}}\equiv ({\vec E}_{n,{\vec k}}, {\vec H}_{n,{\vec k}})^T$ (including both the electric field ${\vec E}$ and the magnetic field ${\vec H}$). Hence for the $k_x=\pi$ plane. Hence, for all Bloch states
\be
\Theta_x^2= e^{i k_x} .
\ee
Thus on the $k_x=\pi$ plane we have $\Theta_x^2=-1$ which explains the double degeneracy on the $k_x=\pi$ plane. Specifically, there exist the following commutation
  relations between anti-unitary operator $\Theta_{x}$ and operator $M_x$ for the $k_x=\pi$ plane,
\be
M_x\Theta_x = - \Theta_x M_x .
\ee
Thus for an eigenstate of $M_x$ with eigenvalue $m_x$, labeled as
$\ket{m_x}$, we have
\be
M_x\Theta_x\ket{m_x} = - \Theta_x M_x\ket{m_x} = - m_x \Theta_x
\ket{m_x} .
\ee
Hence $\Theta_x\ket{m_x}$ is also an eigenstate of $M_x$ with opposite eigenvalue. These are the two degenerate Bloch states on the $k_x=\pi$ plane. 

In a similar way, the anti-unitary operator $\Theta_y$ is a symmetry operator on the $k_y=\pi$ plane. The square of this operator $\left(\Theta_{y}\right)^2: (x,y,z,t) \rightarrow (x,y+1,z,t)$ yields 
\be
\Theta_y^2 = e^{i k_y} 
\ee
for all Bloch states. Thus on the $k_y=\pi$ plane we have $\Theta_y^2 =-1$, which yields the Kramers degeneracy.

Moreover, we find that $\Theta_{gx}\equiv G_x*{\cal T} :
(x, y, z, t) \to (-x, y + \frac{1}{2}, z + \frac{1}{2}, -t)$ and
$\Theta_{gy}\equiv G_y*{\cal T} : (x, y, z, t) \to (x + \frac{1}{2},
-y, z + \frac{1}{2}, -t)$ can also 
yield Kramers degeneracy. Since $\Theta_{gx}$ transforms ${\vec k}$
into $(k_x,-k_y,-k_z)$, it is a symmetry operator only for the four
lines: $k_y, k_z=0, \pi$. The square of this operator $\left(\Theta_{y}\right)^2: (x,y,z,t) \rightarrow (x,y+1,z+1,t)$ yields 
\be
\Theta_{gx}^2=e^{i(k_y+k_z)}
\ee
for all Bloch states. Therefore it leads to double degeneracy for the
two lines: $(k_x, 0, \pi)$ and $(k_x, \pi, 0)$. Similarly,
$\Theta_{gy}$ is a symmetry operator for the four lines: $k_x,
k_z=0,\pi$. The square of this operator $\left(\Theta_{y}\right)^2: (x,y,z,t) \rightarrow (x,y+1,z+1,t)$ yields  
\be
\Theta_{gy}^2=e^{i(k_x+k_z)}
\ee
which reuslt in Kramers degeneracy on the two lines: $(0, k_y, \pi)$ and $(\pi, k_y,
0)$. Notice that these lines crossing at the $Z$ point $(0,0,\pi)$,
where both $\Theta_{gx}$ and $\Theta_{gy}$ are effective.

\section{Sec.~D Proof of the fourfold degeneracy on the $M$-$A$ line}
The fourfold degeneracy on the $M$-$A$ line can be understood 
as follows: Any Bloch state on this line can be labeled with the
eigenvalues $m_x$ and $m_y$ of the two mirror operators $M_x$ and
$M_y$, respectively. We shall prove that $\ket{m_x,m_y}$,
$\Theta_z\ket{m_x,m_y}$ ($\Theta_z\equiv G_z*{\cal T}$),
$S_\pi\ket{m_x,m_y}$, and $\Theta_zS_\pi\ket{m_x,m_y}$ are distinct
from each other. We find that such degeneracy is essentially related
to the following commutation relationships on the $M$-$A$ line,
\begin{align}
& [M_x, \Theta_z]_{+} = 0, \quad [M_y, \Theta_z]_{-}=0,\nn\\
& [M_x, S_\pi]_{+} = 0 ,\quad  [M_y, S_\pi]_{+} = 0 ,
\end{align}
where $[A, B]_{\pm} = AB \pm BA$. 
Thus for an eigenstate labeled with $m_x$ and $m_y$ of mirror operator $M_x$ and $M_y$, we have
\begin{align}
 & M_x \Theta_z \ket{m_x,m_y}=-\Theta_z M_x\ket{m_x,m_y}=-m_x \Theta_z \ket{m_x,m_y}, \nn\\
 & M_y \Theta_z \ket{m_x,m_y}=\Theta_z M_y\ket{m_x,m_y}=m_y \Theta_z \ket{m_x,m_y}, \nn\\
 & M_x S_{\pi} \ket{m_x,m_y}=-S_{\pi} M_x\ket{m_x,m_y}=-m_x S_{\pi} \ket{m_x,m_y}, \nn \\
 & M_y S_{\pi} \ket{m_x,m_y}=-S_{\pi} M_y\ket{m_x,m_y}=-m_y S_{\pi} \ket{m_x,m_y}.
\end{align}
From these relations, we find that: (i) $\Theta_z \ket{m_x,m_y}$ is also an eigenstates of $M_x$ with opposite eigenvalue, i.e.,$\Theta_z\ket{m_x,m_y} = \ket{-m_x,m_y}$; (ii)  $S_{\pi} \ket{m_x,m_y}$ is also an eigenstates of $M_x (M_y)$ with opposite eigenvalue, i.e., $S_\pi\ket{m_x,m_y}=\ket{-m_x,-m_y}$. (ii) $\Theta_z S_\pi$ is also an eigenstates of $M_y$ with opposite eigenvalue, i.e., $ \Theta_z S_\pi\ket{m_x,m_y} = \ket{m_x,-m_y}$. It is evident that these four states, i.e.,  $\ket{m_x,m_y}$,
$\Theta_z\ket{m_x,m_y}$, $S_\pi\ket{m_x,m_y}$, and $\Theta_zS_\pi\ket{m_x,m_y}$ are distinct from each other and they are related by symmetry operators. Therefore, they are
degenerate states.

\section{Sec.~E Proof of the fourfold quadratic degeneracy on the $Z$ point}
The $S_{\frac{\pi}{2}}$ operator transforms the eigenstate of $M_x$ with the
eigenvalue $m_x$ at ${\vec k}$ to a Bloch state at ${\vec k}^\prime$
as the eigenstate of $M_y$ with the same eigenvalue, i.e.,
\be
\Theta_{\frac{\pi}{2}}M_x\Theta_{-\frac{\pi}{2}}=M_y .
\ee
Thus $\Theta_{\frac{\pi}{2}}$ is a symmetry operator only for the $Z$ and $A$
points. Analogous to the Kramers degeneracy, the above yields
fourfold degeneracy at the $Z$ point. The four degenerate states are
$\ket{\Psi}$, $\Theta_{\frac{\pi}{2}}\ket{\Psi}$,
$(\Theta_{\frac{\pi}{2}})^2\ket{\Psi}$, and $(\Theta_{\frac{\pi}{2}})^3\ket{\Psi}$. 

The four degenerate states can be labeled by the eigenvalue of $M_x$,
$M_y$ and $\tilde{G}_z=e^{i\frac{\pi}{4}}G_z$. We find that 
\begin{align}
& [\Theta_{\frac{\pi}{2}}, \tilde{G}_z]_{-} = 0, \quad [(\Theta_{\frac{\pi}{2}})^2,
\tilde{G}_z]_{+} =0 ,\\
& [(\Theta_{\frac{\pi}{2}})^2, M_x]_{-} = 0, \quad [(\Theta_{\frac{\pi}{2}})^2,
M_y]_{-} = 0 .
\end{align}
Thus $\ket{\Psi}$ and $\Theta_{\frac{\pi}{2}}\ket{\Psi}$ carry distinct mirror
eigenvalues but the same $\tilde{G}_z$ eigenvalue, whereas
$\ket{\Psi}$ and $(\Theta_{\frac{\pi}{2}})^2\ket{\Psi}$ carry the same mirror
eigenvalues but different $\tilde{G}_z$ eigenvalues. 
For simplify, we label the Bloch states, which is the eigenstates of $M_x$, $M_y$ and $G_z$ with eigenvalue $m_x$, $m_y$, and $g_z$, respectively, as $\ket{m_x,m_y,g_z}$. According to the above commute (anticommute) relationships, we find that
$\Theta_{\frac{\pi}{2}}\ket{\Psi}=\ket{m_y,m_x,g_z}$, 
$(\Theta_{\frac{\pi}{2}})^2\ket{\Psi}=\ket{m_x,m_y,-g_z}$, and
$(\Theta_{\frac{\pi}{2}})^3\ket{\Psi}=\ket{m_y,m_x,-g_z}$. It is obviously that these four states, i.e., $\ket{m_x,m_y,g_z}$,$\ket{m_y,m_x,g_z}$,$\ket{m_x,m_y,-g_z}$, and $\ket{m_y,m_x,-g_z}$ are distinct from each other and they are related by symmetry operations. Therefore, we prove the fourfold degeneracy at $Z$ point. 
 
\section{Sec.~F Varying geometries and materials}
Here we show the emergent Dirac physics are robust to the material and geometry of the photonic crystals. We calculate the photonic bands along the high symmetry lines with three different geometric/material parameter settings. The results of the photonic bands are shown in Fig.6, where the Dirac nodal line and the quadratic point holds for all parameters. This is because such double degeneracy are guranteed the lattice symmetry. This also indicates the stableness of the Dirac points and implies that such a symmetry-guided method can also be effective in other classicl/bosonic systems.
\begin{figure*}
  \includegraphics[width=15cm]{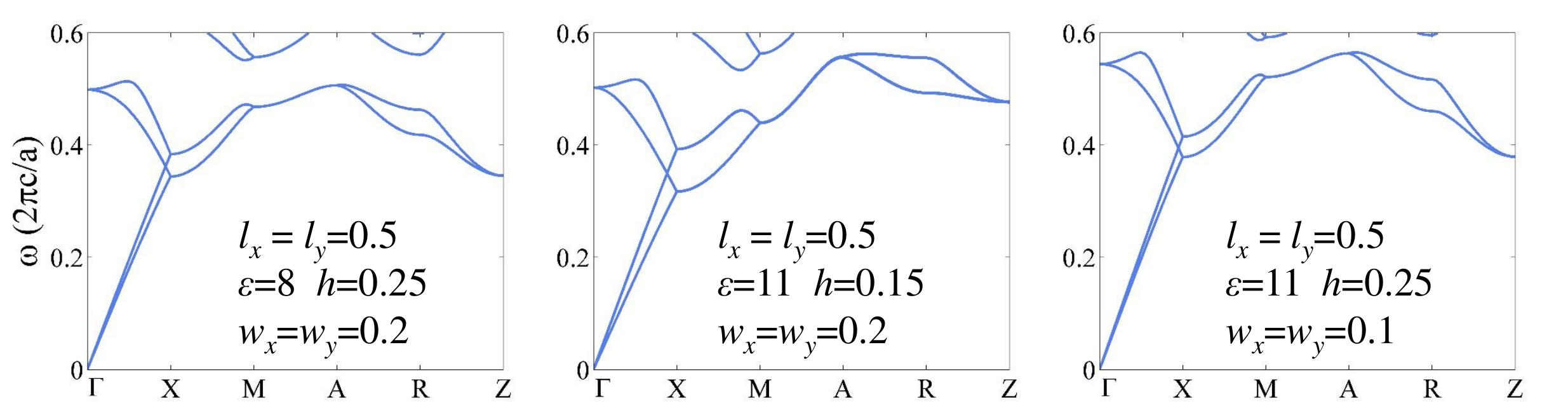}
  \caption{(Color online) Photonic bands along the high symmetry lines for various geometry/material parameters. (a) $l_x=l_y=0.5,w_x=w_y=0.2,h=0.25,\epsilon=8$,  (b) $l_x=l_y=0.5,w_x=w_y=0.2,h=0.15,\epsilon=11$, (c) $l_x=l_y=0.5,w_x=w_y=0.1,h=0.25,\epsilon=11$. } 
\end{figure*}

\section{Sec.~G ${\vec k}\cdot {\vec P}$ Hamiltonian of the Dirac nodal line}
The photonic Hamiltonian is obtained by applying the ${\vec k}\cdot {\vec P}$
method to the Maxwell equation,
\be
\grad\times\frac{1}{\vep}\grad\times {\vec H}_{n^\prime,{\vec k}} =
\frac{\ome_{n^\prime,{\vec k}}^2}{c^2}  {\vec H}_{n^\prime,{\vec k}}  .
\ee
The photonic Hamiltonian $\hat{{\cal H}}_{EM}\equiv
c^2 \grad\times\frac{1}{\vep}\grad\times$ around a high symmetry
${\vec K}$ is obtained as follows. We first expand the Bloch states at
a wavevector ${\vec k}$ in the basis formed by the four Bloch states at
the high symmetry point, i.e., (${\vec q}\equiv {\vec k}-{\vec K}$)
\be
{\vec H}_{n^\prime, {\vec q}} = \sum_{n} e^{i{\vec q}\cdot{\vec r}}
C_n {\vec H}_{n^\prime} ,
\ee
where ${\vec H}_{n^\prime}$ is the magnetic field at the ${\vec q}=0$
point (i.e., the high symmetry point) and $C_n$ are the
coefficients to be solved by diagonalizing the Hamiltonian (and
normalized as $\sum_n|C_n|^2=1)$. The
magnetic fields are normalized such that (UC stands for unit-cell)
\be
\int_{UC} d{\vec r} {\vec H}_{n}^\ast \cdot {\vec H}_{n^\prime} = \delta_{n,n^\prime} .
\ee
The Hamiltiona is written in the basis of the Bloch states ${\vec
  H}_{n^\prime}$, and we find that 
\be
\hat{{\cal H}}_{EM} = \ome_0^2\delta_{n,n^\prime} +\sum_{\alpha} q_\alpha
P_{n,n^\prime}^\alpha + \sum_{\alpha,\beta} W_{n\alpha,n^\prime\beta}q_\alpha
q_\beta .
\ee
Direct calculation yields,
\begin{align}
& P_{n,n^\prime}^\alpha = i \sum_\nu \int_{u.c.}\frac{d{\vec
    r}}{\vep({\vec r})} [H^\ast_{n,\nu} \partial_\nu H_{n^\prime,\alpha}
- H_{n^\prime,\nu} \partial_\nu H_{n,\alpha}^\ast \nn\\
& \quad\quad \quad\quad -H^\ast_{n,\nu} \partial_\alpha H_{n^\prime,\nu} + H_{n^\prime,\nu}  \partial_\alpha
H^\ast_{n,\nu} ] .
\end{align}
We shall use the Maxwell equation to simplify the above results
\be
\partial_\nu H_{n,\alpha} = \sum_\mu - i \frac{\ome_n}{c}
E_{n,\mu}\ep_{\nu\alpha\mu} \vep({\vec r}), \label{aaa}
\ee
where $\ome_n=\ome_0$ at the degenerate point, $\ep_{\nu\alpha\mu}$ is
the Levi-Civita tensor, ${\vec E}_{n}$ is the
electric field of the Bloch states at ${\vec q}=0$ satisfying the normalization condition of
\be
\int_{u.c.} d{\vec r} \vep({\vec r}) {\vec E}_{n}^\ast \cdot {\vec
  E}_{n^\prime} = \delta_{n,n^\prime} .
\ee
Using Eq.~(\ref{aaa}), we find that 
\be
P_{n,n^\prime}^\alpha = 2\ome_0 c\int_{u.c.}d{\vec r} [{\vec E}_{n^\prime}\times
  {\vec H}_{n}^\ast + {\vec E}_{n}^\ast \times {\vec H}_{n^\prime}]\cdot {\vec n}_\alpha
\ee
where $\alpha=(x,y,z)$ and ${\vec n}_{\alpha}$ is the unit vector
along the $\alpha$ direction. And 
\be
W_{n\alpha,n^\prime\beta} = c^2 \int_{u.c.} \frac{d{\vec r}}{\vep({\vec r})}
[\delta_{\alpha\beta} ({\vec H}_n^\ast\cdot{\vec H}_{n^\prime}) -
H_{n\alpha}^\ast H_{n^\prime\beta}] .
\ee
The photonic Hamiltonian $\hat{{\cal H}}_{EM}$ is connected with the
simulated fermion-like as follows, ($\hbar\equiv 1$)
\be
\hat{{\cal H}}_{EM} := (\hat{{\cal H}}_F)^2 .
\ee
Therefore, near the degenerate point, we have 
\be
\hat{{\cal H}}_F = \ome_0 + \sum_\alpha \hat{v}^\alpha q_\alpha + \sum_{\alpha,\beta} \hat{w}_{\alpha,\beta}q_\alpha
q_\beta  + ...
\ee
where $...$ represents higher order terms, and
\begin{align}
& \hat{v}^\alpha = \frac{1}{2\ome_0} P_{n,n^\prime}^\alpha ,\nn\\
& \hat{w}_{\alpha,\beta} = \frac{1}{2\ome_0} \hat{W}_{\alpha,\beta} .
\end{align}
One can easily verify that $\hat{v}^\alpha$ and
$\hat{w}_{\alpha,\beta}$ are Hermitian operators.

We now examine the constraints on the matrix element of $\hat{v}$
imposed by the mirror and/or glide symmetry. Explicitly, we have
\be
v_{n,n^\prime}^\alpha = c \sum_{\mu,\nu} \ep_{\alpha \mu\nu}
\int_{u.c.}d{\vec r} [ E_{n^\prime,\mu} H_{n,\nu}^\ast +
E_{n,\mu}^\ast H_{n^\prime,\nu}] .
\ee
If, say, there is a mirror or glide symmetry, labeled as
$\hat{F}_{x}$, the mirror operation acts on the 
electric and magnetic fields as follows:
\begin{align}
& \hat{F}_x H_x({\vec r}) = H_x(\hat{F}_x{\vec r})  ,\nn \\
& \hat{F}_x H_y({\vec r}) = - H_y(\hat{F}_x{\vec r}) ,\nn \\
& \hat{F}_x H_z({\vec r}) = - H_z(\hat{F}_x{\vec r}) ,\nn \\
& \hat{F}_x E_x({\vec r}) = - E_x(\hat{F}_x{\vec r})  ,\nn \\
& \hat{F}_x E_y({\vec r}) = E_y(\hat{F}_x{\vec r}) ,\nn \\
& \hat{F}_x E_z({\vec r}) = E_z(\hat{F}_x{\vec r})  .
\end{align}
Thus, the velocity matrix element $v_{n,n^\prime}^\alpha$ is nonzero
only when the $n$ and $n^\prime$ states carry opposite mirror
eigenvalue $m_\alpha$ (or glide eigenvalue $g_\alpha$).

The invariance of the Hamiltonian under a symmetry operation ${\cal S}$ implies that
\be
{\cal S}  {\cal H}({\vec k}) {\cal S}^{-1} = {\cal H}({\cal S}{\vec
  k}) . \label{sym}
\ee
For example, the time-reversal symmetry symmetry ${\cal T}=-{\cal K}$
yields that the ${\vec k}\cdot{\vec P}$ around a time-reversal
invariant momentum has
\be
(v_{n,n^\prime}^\alpha )^\ast = - v_{n,n^\prime}^\alpha
\ee
Hence the matrix element of the $q$ linear terms are purely imaginary.
One can prove that the $q$ quadratic terms are purely real.

We now derive the ${\vec k}\cdot{\vec P}$ Hamiltonian for the Dirac
nodal line. Around a point on the $M$-$A$ line, the four degenerate states
are chosen as the $\ket{m_x,m_y}$ for $m_x,m_y=\pm 1$. In the basis of
$(\ket{1,1}, \ket{-1,-1}, \ket{-1,1}, \ket{1,-1})^T$ the $q$
linear Hamiltonian is written as
\be
\hat{{\cal H}}_F^{DL} =\ome_0 + \left( \begin{array}{cccccccccccc}
    0 & 0 & a_1 q_x & a_2 q_y \\
    0 & 0 & b_1 q_y & b_2 q_x  \\
    a_1^\ast q_x & b_1^\ast q_y & 0 & 0 \\
    a_2^\ast q_y & b_2^\ast q_x & 0 & 0 \\
    \end{array}\right) , \label{hdl}
\ee
where $a_i$ and $b_i$ are the ${\vec k}\cdot{\vec P}$ coefficients
which can be calculated from the photonic Bloch functions ${\vec
  H}_{n}$ at the degeneracy point $(\pi,\pi,k_z)$ (hence these
coefficients are $k_z$ dependent). Those coefficients are restricted
by the symmetry via Eq.~(\ref{sym}). Two symmetry operations are
relevant: $\Theta_z$ and $S_\pi$. In the chosen basis, these
operations are manifested as the following matrices:
\be
\Theta_z = \tau_y {\cal K}, \quad S_\pi = \sigma_y e^{ik_z z/2} .
\ee
Their effects on the wavevectors are
\be
\Theta_z {\vec k} = ( -k_x, -k_y, k_z), \quad S_\pi {\vec k} = ( -
k_x, - k_y, k_z) .
\ee
We thus obtain from Eq.~(\ref{sym}) that
\be
b_1 = a_2, \quad b_2 = - a_1 .
\ee
It is convenient to define
$v=\sqrt{|b_1|^2+|b_2|^2}=\sqrt{|a_1|^2+|a_2|^2}$, $\gamma_x=a_1/v$, and
$\gamma_y=a_2/v$. In the new basis of $(\ket{\up,p}, \ket{\down,p},
\ket{\up,a}, \ket{\down,a})^T$, the ${\vec k}\cdot{\vec P}$
Hamiltonian can be further simplified as 
\begin{align}
& \hat{{\cal H}}_F^{DL} = \ome_0 + v 
\left( \begin{array}{ccccc}
    0 & \hat{{\cal A}} \\
    \hat{{\cal A}}^\dagger &  0 \\
  \end{array}\right) + {\cal O}(q^2) , \nn\\
& \hat{{\cal A}} \equiv \gamma_x q_x \sigma_x + \gamma_y q_y \sigma_y .
\end{align}
The above Hamiltonian applies for the whole $M$-$A$ line, where the
coefficients $\ome_0$, $v$, $\gamma_x$ and $\gamma_y$ are
$k_z$-dependent. At the M or A point, the $S_4$ symmetry also holds,
thus there will be additional constraints on the coefficients.
Let's consider such constraints in the original basis for the
Hamiltonian Eq.~(\ref{hdl}). $S_4$ is manifested as $-\sigma_0$ in
the even-parity doublet, whereas $S_4=-i\sigma_y$ in the odd-parity
doublet. Imposing (\ref{sym}) we find that for the M and A points,
$\gamma_x=\gamma_y$. In addition, for these points, the time-reversal
symmetry dictates that $\gamma_x=\gamma_y$ are purely imaginary
coefficients. Therefore, at these points the Dirac point is
doubly degenerate in the $k_x$-$k_y$ plane, while away from these
points the dispersion in the $k_x$-$k_y$ plane is generally
non-degenerate. The double degeneracy for generic points on the MA
line (except the M and A points) are restored only for $q_x=0$ or
$q_y=0$ (i.e., at the $k_x=\pi$ or $k_y=\pi$ plane), where the
nonsymmorphic screw symmetry ensures double degeneracy.
 
For the quadratic degeneracy at the $Z$ point, the eigenstates can be
labeled as $\ket{m_x,m_y,g_z}$ for $m_x=-m_y=\pm1, g_z=\pm 1$. Using
the basis of $(\ket{-1,1,1}, \ket{1,-1,1}, \ket{-1,1,-1},
\ket{1,-1,-1})^T$, the $q$ linear term comes simply as $v_z
q_z\tau_y$, because it is finite only between states with opposite
$g_z$. Considering the $S_{\frac{\pi}{2}}$ symmetry, the higher order
terms are 
\begin{widetext}
\be
f_0 f_2 q_{\parallel}^2 + f_0 \left( \begin{array}{cccccccccccc}
    q_x^2 - q_y^2 & 2 f_1 q_x q_y & 0 & 0 \\
    2f_1 q_x q_y & q_y^2-q_x^2 &  0 & 0  \\
    0 & 0 & q_x^2 - q_y^2 & 2f_1 q_x q_y \\
    0 & 0 & 2f_1 q_x q_y & q_y^2 - q_x^2 \\
    \end{array}\right) . \label{hz}
\ee
\end{widetext}
The above Hamiltonian recovers Eq.~(8) in the main text when written
in the basis of spin states that carrying angular momentum. When the
$S_{\frac{\pi}{2}}$ symmetry is broken the degeneracy between states
with different $m_x$ at $q_z=0$ is split by a constant term
$\Delta_zf_0\sigma_z$ with $\Delta_z$ characterizing the strength of the
perturbation.

\section{Sec.~H Optical properties of type-II Dirac points}
The type-II DPs offer special band structures that may enable in the manipulation of light in ways that cannot be achieved in uniform dielectric materials. The refraction properties can be
determined by matching the frequency and the wavevector parallel to
the interface. Here, we consider refraction on the $(001)$ surface
where light is injected from air (above the PC). The wavevector in the
air is determined by the frequency and angle of incidence ($\theta_i$ and $\phi_i$) via
\begin{align}
& k_x = \frac{\ome}{c} \sin\theta_i\cos\phi_i , \quad k_y =
\frac{\ome}{c} \sin\theta_i\sin\phi_i ,\nn\\
& k_z = \frac{\ome}{c} \cos\theta_i .
\end{align}
The wavevector $k_x$ and $k_y$ remains the same in the PC, the
quantity to be found is the wavevector along $z$ direction in the
PC. We shall denote the wavevector in the PC as ${\vec q}$. So far we
have 
\be
q_x = k_x, \quad q_y = k_y .
\ee
For the Dirac points, $q_z$ is obtained by solving the equation,
\begin{widetext}
\begin{align}
& \delta \ome = v_z q_z\tau_z + f_0 \Bigg[\beta
q_\parallel^2 \pm \sqrt{(\Delta_z +q_x^2 -q_y^2)^2 + 4 q_x^2
  q_y^2  }\bigg] ,\nn
\end{align}
for the $\pm$ branches with $\tau_z=-1$ as required by that the group
velocity along the $z$ direction should be negative. The above
equation can be solved straightforwardly.

The group velocity along the $x$ and $y$ directions are then,
\begin{align}
& v_{g,x} = \frac{\partial\ome}{\partial q_x} = f_0 \bigg(2\beta
q_x \pm \frac{2(\Delta_z + q_x^2 -q_y^2) q_x + 4 q_y^2 q_x}{\sqrt{(\Delta_z +q_x^2 -q_y^2)^2 + 4 q_x^2 q_y^2 } }\bigg) , \nn\\
& v_{g,y} = \frac{\partial\ome}{\partial q_y} = f_0 \bigg(2\beta
q_y \pm \frac{-2(\Delta_z + q_x^2 -q_y^2) q_y + 4 q_x^2 q_y}{\sqrt{(\Delta_z +q_x^2 -q_y^2)^2 + 4 q_x^2 q_y^2 } }\bigg) .
\end{align}
The refraction angle $\phi_o$ is then determined as
\be
\phi_o = {\rm Arg}[v_{g,x} + i v_{g,y}] 
\ee
for the $\pm$ branches. The crucial physics is that around the type-II
Dirac point the $v_{g,x}$ becomes significant for the two branches and
they are of opposite sign, while the $v_{g,y}$ do not change
significantly. This yields a large variation of the refraction angle
across the Dirac point, such variation goes to opposite direction
for the two branches.
\end{widetext}


\begin{thebibliography}{999}

\bibitem{volovik} G. E. Volovik, {\it The Universe in a Helium Droplet},
\href{https://oxford.universitypressscholarship.com/view/10.1093/acprof:oso/9780199564842.001.0001/acprof-9780199564842?rskey=ZFiUYw&result=1}{(Oxford: Clarendon Press, 2003)}.


\bibitem{wan} X. Wan, A. M. Turner, A. Vishwanath, and S. Y. Savrasov, {\it Topological semimetal and Fermi-arc surface states in the electronic structure of pyrochlore iridates}, \href{https://journals.aps.org/prb/abstract/10.1103/PhysRevB.83.205101}{Phys. Rev. B {\bf 83}, 205101 (2011)}.


\bibitem{fang} Z. Wang, Y. Sun, X.-Q. Chen, C. Franchini,
  G. Xu, H. Weng, X. Dai, and Z. Fang, {\it Dirac semimetal and topological phase transitions in ${A}_{3}$Bi ($A=\text{Na}$,    K, Rb)}, \href{https://journals.aps.org/prb/abstract/10.1103/PhysRevB.85.195320}{Phys. Rev. B {\bf 85}, 195320 (2012)}.


\bibitem{liu} Z. K. Liu, B. Zhou, Y. Zhang, Z. J. Wang, H. M. Weng,
D. Prabhakaran, S. K. Mo, Z. X. Shen, Z. Fang, X. Dai, Z. Hussain,
and Y. L. Chen, {\it Discovery of a three dimensional topological Dirac semimetal Na$_3$Bi}, \href{https://science.sciencemag.org/content/343/6173/864.abstract}{Science {\bf 343}, 864 (2014)}.


\bibitem{WPII} A. A. Soluyanov,	D. Gresch, Z. Wang, Q. Wu, M. Troyer, X. Dai, and B. A. Bernevig, {\it Type-II Weyl semimetals}, \href{https://www.nature.com/articles/nature15768}{Nature (London) {\bf 527}, 495 (2015)}.


\bibitem{science} B. Bradlyn, J. Cano, Z. Wang, M. G. Vergniory, C. Felser, R. J. Cava, and B. A. Bernevig, {\it Beyond Dirac and Weyl fermions: unconventional quasiparticles in conventional crystals}, \href{https://www.science.org/doi/10.1126/science.aaf5037}{Science {\bf 353}, aaf5037 (2016)}. 


\bibitem{mele} N. P. Armitage, E. J. Mele, and Ashvin Vishwanath, {\it Weyl and Dirac semimetals in three-dimensional solids}, Rev. Mod. Phys. {\bf 90}, 015001 (2018).

\bibitem{ling1} L. Lu, L. Fu, J. D. Joannopoulos, and M. Solja\v{c}i\'{c}, {\it Weyl points and line nodes in gyroid 	photonic crystals}, \href{https://www.nature.com/articles/nphoton.2013.42}{Nat. Photon. {\bf 7}, 294 (2013)}.


\bibitem{ling-exp} L. Lu, Z. Wang, D. Ye, L. Ran, L. Fu, J. D. Joannopoulos, and M. Solja\v{c}i\'{c}, {\it Experimental observation of Weyl points},
\href{https://science.sciencemag.org/content/349/6248/622}{Science {\bf 349}, 622 (2015)}.


\bibitem{szhang1} W. Gao, B. Yang, M. Lawrence, F. Fang, B. B\'eri, and S. Zhang, {\it Plasmon Weyl degeneracies in magnetized plasma},
\href{https://www.nature.com/articles/ncomms12435}{Nat. Comm. {\bf 7}, 12435 (2016)}.


\bibitem{SZhang} D. Wang {\sl et al}., {\it Photonic Weyl points due to broken time-reversal symmetry in magnetized semiconductor}, Nat. Phys. {\bf 15}, 1150-1155 (2019).

  
  
\bibitem{ct-exp} W.-J. Chen, M. Xiao, and C. T. Chan, {\it Photonic crystals possessing multiple Weyl points and the experimental observation of robust surface states}, \href{https://www.nature.com/articles/ncomms13038}{Nat. Commun. {\bf 7}, 13038 (2016)}.


\bibitem{3ddp} H.-X. Wang, L. Xu, H.Y. Chen, and J.-H. Jiang, {\it 	Three-dimensional photonic Dirac points stabilized by point group 	symmetry}, \href{https://journals.aps.org/prb/abstract/10.1103/PhysRevB.93.235155}{Phys. Rev. B {\bf 93}, 235155 (2016)}.


\bibitem{xiao} M. Xiao, Q. Lin, and S. Fan, {\it Hyperbolic Weyl point in reciprocal chiral metamaterials}, \href{https://journals.aps.org/prl/abstract/10.1103/PhysRevLett.117.057401}{Phys. Rev. Lett. {\bf 117}, 057401 (2016)}.
    
    
\bibitem{sWP} Q. Lin, M. Xiao, L. Yuan, and S. Fan, {\it Photonic Weyl point in a two-dimensional resonator lattice with a synthetic frequency dimension}, \href{https://www.nature.com/articles/ncomms13731}{Nat. Commun. {\bf 7}, 13731 (2016)}.


\bibitem{atwater} S. Peng, R. Zhang, V. H. Chen, E. T. Khabiboulline, P. Braun, and H. A. Atwater, {\it Three-Dimensional Single Gyroid Photonic Crystals with a Mid-Infrared Bandgap}, \href{https://pubs.acs.org/doi/abs/10.1021/acsphotonics.6b00293}{ACS Photon. {\bf 3}, 1131 (2016)}.


\bibitem{type2} H.-X. Wang, Y. Chen, Z. H. Hang, H.-Y. Kee, and J.-H. Jiang, {\it Type-II Dirac photons},
\href{https://www.nature.com/articles/s41535-017-0058-z}{npj Quantum Mater.  {\bf 2}, 54 (2017)}.

  
\bibitem{Rechtsman1} J. Noh, S. Huang, D. Leykam, Y. D. Chong, K. P. Chen, and M. C. Rechtsman, {\it Experimental observation of optical Weyl points and Fermi arc-like surface states}, \href{http://dx.doi.org/10.1038/nphys4072}{Nat. Phys. {\bf 13}, 611 (2017)}.


\bibitem{ExptypeIWeyl} E. Goi, Z. Yue, B. P. Cumming, and M. Gu, {\it Observation of type-I Photonic Weyl points in optical frequencies}, \href{ https://doi.org/10.1002/lpor.201700271}{Laser \& Photon. Rev. {\bf 12} 1700271 (2018)}.
 
 
\bibitem{Rechtsman2} S. Vaidva, J. Noh, A. Cerian, C. J{\"o}rg, G. v. Freymann, and M. C. Rechtsman, {\it Observation of a charge-2 photonic Weyl point in the infrared}, \href{https://journals.aps.org/prl/abstract/10.1103/PhysRevLett.125.253902}{Phys. Rev. Lett. {\bf 125}, 253902 (2020)}.
  
  
\bibitem{Rechtsman3} C. J{\"o}rg, S. Vaidva, J. Noh, A. Cerian, S. Augustine, G. v. Freymann, and M. C. Rechtsman, {\it Observation of the spliting of charge-2 (quadratic) Weyl points in near-infrared photonic crystals}, \href{https://arxiv.org/abs/2106.12119v2}{arXiv: 2106.12119 (2021)}.
  

\bibitem{szhang2} Q. Guo, B. Yang, L. Xia, W. Gao, H. Liu, J. Chen, Y. Xiang, and S. Zhang, {\it Three dimensional photonic Dirac points in metamaterials}, \href{https://journals.aps.org/prl/abstract/10.1103/PhysRevLett.119.213901}{Phys. Rev. Lett. {\bf 119}, 213901 (2017)}.


\bibitem{ctwood} M.-L. Chang, M. Xiao, W.-J. Chen, and C. T. Chan,
{\it Multi Weyl points and the sign change of their	topological charges in woodpile photonic crystals}, \href{https://journals.aps.org/prb/abstract/10.1103/PhysRevB.95.125136}{Phys. Rev. B {\bf 95}, 125136 (2017)}.


\bibitem{ZB} X. Zhang, {\it Observing Zitterbewegung for Photons near the Dirac Point of a Two-Dimensional Photonic Crystal}, \href{https://journals.aps.org/prl/abstract/10.1103/PhysRevLett.100.113903}{Phys. Rev. Lett. {\bf 100}, 113903 (2008)}.


\bibitem{psd} R. A. Sepkhanov, Ya. B. Bazaliy, and C. W. J. Beenakker, {\it Extremal transmission at the Dirac point of a photonic band structure}, \href{https://journals.aps.org/pra/abstract/10.1103/PhysRevA.75.063813}{Phys. Rev. A {\bf 75}, 063813 (2007)}.


\bibitem{zim} X. Q. Huang, Y. Lai, Z. H. Hang, H. H. Zheng, and
C. T. Chan, {\it Dirac cones induced by accidental degeneracy in photonic crystals and zero-refractive-index materials}, \href{https://www.nature.com/articles/nmat3030}{Nat. Mater. {\bf 10}, 582 (2011)}.


\bibitem{sMag} M. C. Rechtsman, J. M. Zeuner, A. T\"unnermann,
S. Nolte, and M. Segev, and A. Szameit, {\it Strain-induced pseudomagnetic field and photonic Landau levels in dielectric structures}, \href{https://www.nature.com/articles/nphoton.2012.302}{Nat. Photon. {\bf 7}, 153 (2013)}.


\bibitem{haldane} F. D. M. Haldane and S. Raghu, {\it Possible Realization of Directional Optical Waveguides in Photonic Crystals with Broken Time-Reversal Symmetry}, Phys. Rev. Lett. {\bf 100}, 013904 (2008) 


\bibitem{wangzhen} Z. Wang, Y. D. Chong, J. D. Joannopoulos, and M. Solja\v{c}i\'{c}, {\it Reflection-Free One-Way Edge Modes in a Gyromagnetic Photonic Crystal}, Phys. Rev. Lett. {\bf 100}, 013905 (2008)


\bibitem{CI1} Z. Wang, Y. Chong, J. D. Joannopoulos, and M. Solja\v{c}i\'{c}, {\it Observation of unidirectional backscattering immune topological electromagnetic states}, \href{https://www.nature.com/articles/nature08293}{Nature {\bf 461}, 772-775 (2009)}.


\bibitem{hafezi} M. Hafezi, S. Mittal, J. Fan, A. Migdall, and J. M. Taylor, {\it Imaging topological edge states in silicon photonics}, \href{https://www.nature.com/articles/nphoton.2013.274}{
 Nat. Photon. {\bf 7}, 1001-1005 (2013)}.


\bibitem{FTI} M. C. Rechtsman {\sl et al.}, {\it Photonic Floquet topological insulators}, \href{https://www.nature.com/articles/nature12066}{Nature {\bf 496}, 196-200 (2013)}.


\bibitem{z2meta} A. B. Khanikaev, S. H. Mousavi, W.-K. Tse, M. Kargarian, A. H. MacDonald, and G. Shvets, {\it Photonic topological insulators}, \href{https://www.nature.com/articles/nmat3520}{Nat. Mater. {\bf 12}, 233 (2013)}.
  
    
\bibitem{ctti} W.-J. Chen, S.-J. Jiang, X.-D. Chen, J.-W. Dong, and C. T. Chan, {\it Experimental realization of photonic topological insulator in a uniaxial metacrystal waveguide},
\href{https://www.nature.com/articles/ncomms6782}{Nat. Commun. {\bf 5}, 5782 (2014)}.


\bibitem{shvets} T. Ma, A. B. Khanikaev, S. H. Mousavi, and G. Shvets, {\it Guiding electromagnetic waves around sharp corners: topologically protected photonic transport in metawaveguides}, 
\href{https://journals.aps.org/prl/abstract/10.1103/PhysRevLett.114.127401}{Phys. Rev. Lett. {\bf 114}, 127401 (2015)}.


\bibitem{huxiao} L.-H. Wu and X. Hu, {\it Scheme for Achieving a Topological Photonic Crystal by Using Dielectric Material},
\href{https://journals.aps.org/prl/abstract/10.1103/PhysRevLett.114.223901}{Phys. Rev. Lett. {\bf 114}, 223901 (2015)}.


\bibitem{lumh} C. He, X.-C. Sun, X.-P. Liu, M.-H. Lu, Y. Chen,
L. Feng, and Y.-F. Chen, {\it Photonic topological insulator with broken time-reversal symmetry},
\href{https://www.pnas.org/content/113/18/4924.short}{Proc. Natl. Acad. Sci. USA {\bf 113}, 4924 (2016)}.
 

\bibitem{oe1} L. Xu, H.-X. Wang, Y.D. Xu, H.Y. Chen, and J.-H. Jiang, {\it Accidental degeneracy in photonic bands and topological phase transitions in two-dimensional core-shell dielectric photonic crystals}, 
\href{https://www.osapublishing.org/oe/fulltext.cfm?uri=oe-24-16-18059&id=348175}{Opt. Express {\bf 24}, 18059 (2016)}.


\bibitem{3dti} L. Lu, C. Fang, L. Fu, S. G. Johnson, J. D.
Joannopoulos, and M. Solja\v{c}i\'{c}, {\it Symmetry-protected topological photonic crystal in three dimensions}, 
\href{https://www.nature.com/articles/nphys3611}{Nat. Phys. {\bf 12}, 337 (2016)}.


\bibitem{3dwti} A. Slobozhanyuk, S. H. Mousavi, X. Ni, D. Smirnova, Y. S. Kivshar, and A. B. Khanikaev, {\it Three-dimensional all-dielectric photonic topological insulator},
\href{https://www.nature.com/articles/nphoton.2016.253}{Nat. Photon. {\bf 11}, 130 (2017)}.

  
\bibitem{wood1} S. Y. Lin, J. G. Fleming, D. L. Hetherington, B. K. Smith, R. Biswas, K. M. Ho, M. M. Sigalas, W. Zubrzycki, S. R. Kurtz, and J. Bur, {\it A three-dimensional photonic crystal operating at infrared wavelengths}, \href{https://www.nature.com/articles/28343}{Nature (London) {\bf 394}, 251 (1998)}.


\bibitem{wood2} S. Noda, K. Tomoda, N. Yamamoto, and A. Chutinan, {\it Full three-dimensional photonic bandgap crystals at near-Infrared wavelengths}, \href{https://science.sciencemag.org/content/289/5479/604.abstract}{Science {\bf 289}, 604 (2000)}. 


\bibitem{wood3} M. Deubel, G. von Freymann, M. Wegener, S. Pereira,
K. Busch, and C. M. Soukoulis, {\it Direct laser writing of three-dimensional photonic-crystal templates for telecommunications}, \href{https://www.nature.com/articles/nmat1155}{Nat. Mater. {\bf 3}, 444 (2004)}.
  
  
\bibitem{noda} S. Noda, M. Fujita, and T. Asano, {\it Spontaneous-emission control by photonic crystals and nanocavities}, \href{https://www.nature.com/articles/nphoton.2007.141}{Nat. Photon. {\bf 1}, 449-458 (2007)}.


\bibitem{book} J. D. Joannopoulos, S. G. Johnson, J. N. Winn, and
R. D. Meade, {\it Photonic crystals: molding the flow of light},
 \href{http://ab-initio.mit.edu/book/}{(Princeton university press, 2011)}.


\bibitem{wood4} L. A. Ibbotson, A. Demetriadou, S. Croxall, O. Hess, and J. J. Baumberg, {\it Optical nano-woodpiles: large-area metallic photonic crystals and metamaterials}, \href{https://www.nature.com/articles/srep08313}{Sci. Rep. {\bf 5}, 8313 (2015)}.


\bibitem{phc-valley} M. I. Shalaev, W. Walasik, A. Tsukernik, Y. Xu, and N. M. Litchinitser, {\it Robust topologically protected transport in photonic crystals at telecommunication wavelengths}, \href{https://www.nature.com/articles/s41565-018-0297-6}{Nat. Nanotech. {\bf 14}, 31 (2019)}.


\bibitem{Dirac} P. A. M. Dirac, {\it The quantum theory of the electron}, \href{https://royalsocietypublishing.org/doi/abs/10.1098/rspa.1928.0023}{Proc. Roy. Soc. A {\bf 117}, 610-624 (1928)}.

\bibitem{SM} See Supplemental Materials.




\end{thebibliography}
\end{document}